\def   \ni {\noindent}

\def   \ssk {\vskip  5truept}

\def   \bsk {\vskip 15truept}
 
\def   \newpage {\vfill\eject}
\def   \newline {\hfil\break}

\documentstyle[epsfig]{article}
\begin{document}

\hsize 5truein
\vsize 8truein
\font\abstract=cmr8
\font\keywords=cmr8
\font\caption=cmr8
\font\references=cmr8
\font\text=cmr10
\font\affiliation=cmssi10
\font\author=cmss10
\font\mc=cmss8
\font\title=cmssbx10 scaled\magstep2
\font\alcit=cmti7 scaled\magstephalf
\font\alcin=cmr6 
\font\ita=cmti8
\font\mma=cmr8
\def\ref{\par\noindent\hangindent 15pt}
\null


\title{\ni ANALYSIS OF LINE CANDIDATES IN GAMMA-RAY BURSTS OBSERVED BY BATSE}

\bsk \bsk
\author{\ni 
M. S. Briggs$^1$, D. L. Band$^2$, R. D. Preece$^1$,
W. S. Paciesas$^1$ and G. N. Pendleton$^1$ }                                              
\bsk
\affiliation{1) 
Department of Physics, Univ. of Alabama in Huntsville, Huntsville, AL 35899}
\break
\hspace*{4mm}
\affiliation{2) 
Center for Astrophysics and Space Science, Univ. of Calif., San Diego,
CA 92093}
\bsk
\baselineskip = 12pt

\abstract{ABSTRACT. \ni
A comprehensive search of BATSE Spectroscopy Detector data from 117
GRBs has uncovered 13 statistically significant line candidates.
The case of a candidate in GRB~930916 is discussed.
In the data of SD~2 there appears to be a emission line at 46 keV,
however the line is not seen in the data of SD~7.
Simulations indicate that the lack of agreement between the results
from SD~2 and SD~7 is implausible but not impossible.
}                                                    
\bsk
\baselineskip = 12pt

\bsk
\baselineskip = 12pt


\text{\ni 1. INTRODUCTION}
\ssk
\ni

A primary goal of adding the Spectroscopy Detectors (SDs) to 
BATSE was the detection of low-energy spectral
features in gamma-ray bursts.  At the time, the reported low-energy 
lines were
interpreted as resonant cyclotron scattering in intense magnetic fields of
neutron stars in our Galaxy.   
While the former theoretical explanation of spectral features is now untenable
unless there are two populations of GRB sources,
the observational status of spectral features is still important.  

Each of the eight BATSE modules contains one SD, which consists of a 12.7~cm
diameter by 7.6~cm thick crystal of NaI(Tl) viewed by a single
12.7~cm photomultiplier tube.     Compared to the BATSE Large Area Detectors,
the SDs have better energy resolution and a higher probability of full-energy
absorption of incident gamma-rays, but a smaller area.

After the failure of our manual search to
discovery a single line [1], we developed an automatic computer search to
comprehensively search the data of bright bursts [2]. 
Because we do not a priori know the energy, starting time, or duration of
spectral features, the procedure searches a wide range of centroid energies and
timescales.  
Many combinations of consecutive spectra are examined: all
singles, 
pairs, triples, and groups of 4, 5, 7, . . . spectra, up to
the entirety of the high time resolution SD data.
The presence of a line is tested by fitting each spectrum twice, first
with Band's ``GRB'' continuum function,
then with the same continuum function plus a narrow line.
A change in $\chi^2$ of more than
20 identifies a line candidate.   The present search is limited to low-energy
features, so a closely spaced grid of trial centroids extending up to 100 keV is
used.  The LLD is typically just below 20 keV and, after requiring a continuum
interval below the first search centroid, lines are tested starting above 20
keV.

\newpage

The search was applied to 120,700 spectra from 117 GRBs.    Because of the
examination of trial spectra with a sliding starting time and a wide range of
durations, many of these spectra have substantial overlap.  Additionally,
most of the spectra have very low signal-to-noise ratios and consequently a
real spectral feature could not have been detected; these spectra were searched
as controls.  Thus the number of
independent spectra with sufficient photons to support the detection of
a real feature is much lower, below about 1000.


\bsk
{\ni 2. RESULTS}
\ssk
\ni 

The comprehensive search identified 13 candidates.  The $\chi^2$ improvement
from adding a line ranges from 20, the candidate threshold, to 50,
corresponding to chance probabilities of
$4 \times 10^{-5}$ to $10^{-11}$.  The probabilities are calculated for
adding two-parameters (line intensity and centroid; the intrinsic width is
assumed to be narrow) to the spectral fit to a single spectrum.
The energy range searched contains about  five resolution elements;
the number of independent spectra of sufficient intensity searched is below
about 1000. 
Consequently at most one of these candidates might be a chance
fluctuation in the ensemble.

An advantage of BATSE is the observation of bursts by several detectors,
thereby enabling further tests of the
reality of a candidate.  
Is the candidate detected with high statistical
significance in the second detector?  This would be confirmation of the reality
of the feature.    Confirmation might not be achieved for several reasons.  If
the feature is not highly significant in the second detector but a sensitivity
analysis shows that this is reasonable based upon the line strength and the 
viewing
angle of the detector, then the data are consistent.
However, a contradiction obtains
if a sensitivity analysis indicates that the feature
should have been
detected in a second detector but the feature was not detected.

We have previously reported details on two of the candidates.
For GRB~940703 (trigger 3057), only SD~5 had a suitable gain and viewing angle.
A highly significant 44 keV emission line ($\Delta \chi^2$ = 31.2, $P =
2 \times 10^{-7}$) was observed in a portion of that
burst [2].
Because of the gains or viewing angles of the other SDs, no consistency
tests are possible for this candidate.
The other candidate previously described, GRB~941017 (trigger 3245), was
usefully observed by SDs~0 and 5 [3]. 
An apparent emission line at 43 keV was discovered in the data of
SD~0.   A less significant feature appears in SD~5 at a strength consistent with
that feature seen in SD~0. 
This appears to be one of the best cases for detecting a line:
the data from two detectors are consistent and
a joint fit of their data has 
$\Delta \chi^2 = 28.6$  ($P = 6 \times 10^{-7}$) for adding a line.

\begin{figure}[tb!]
\centerline{
\mbox{
\epsfig{file=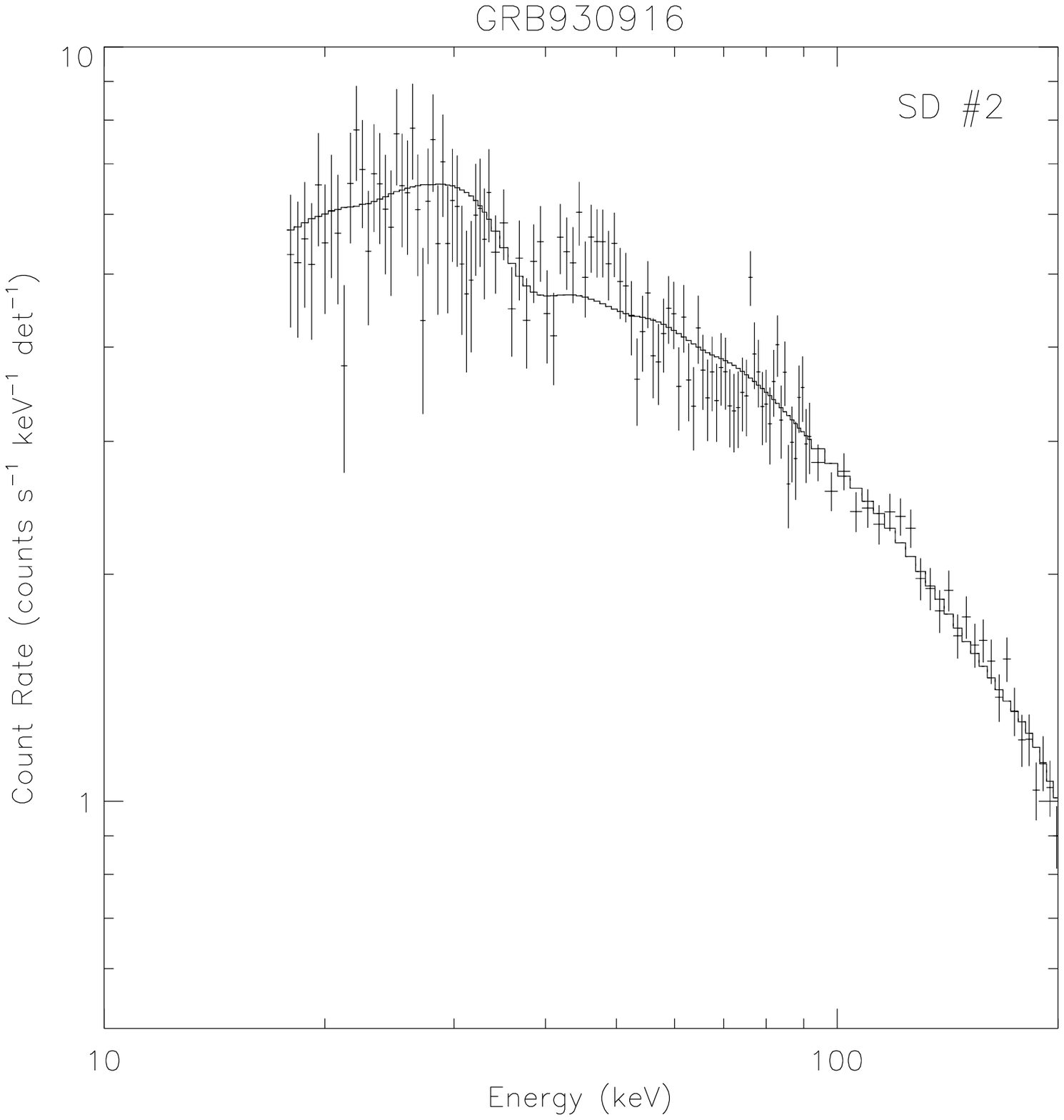,width=6.0cm,%
bbllx=55bp,bblly=185bp,bburx=550bp,bbury=674bp,clip=}
\epsfig{file=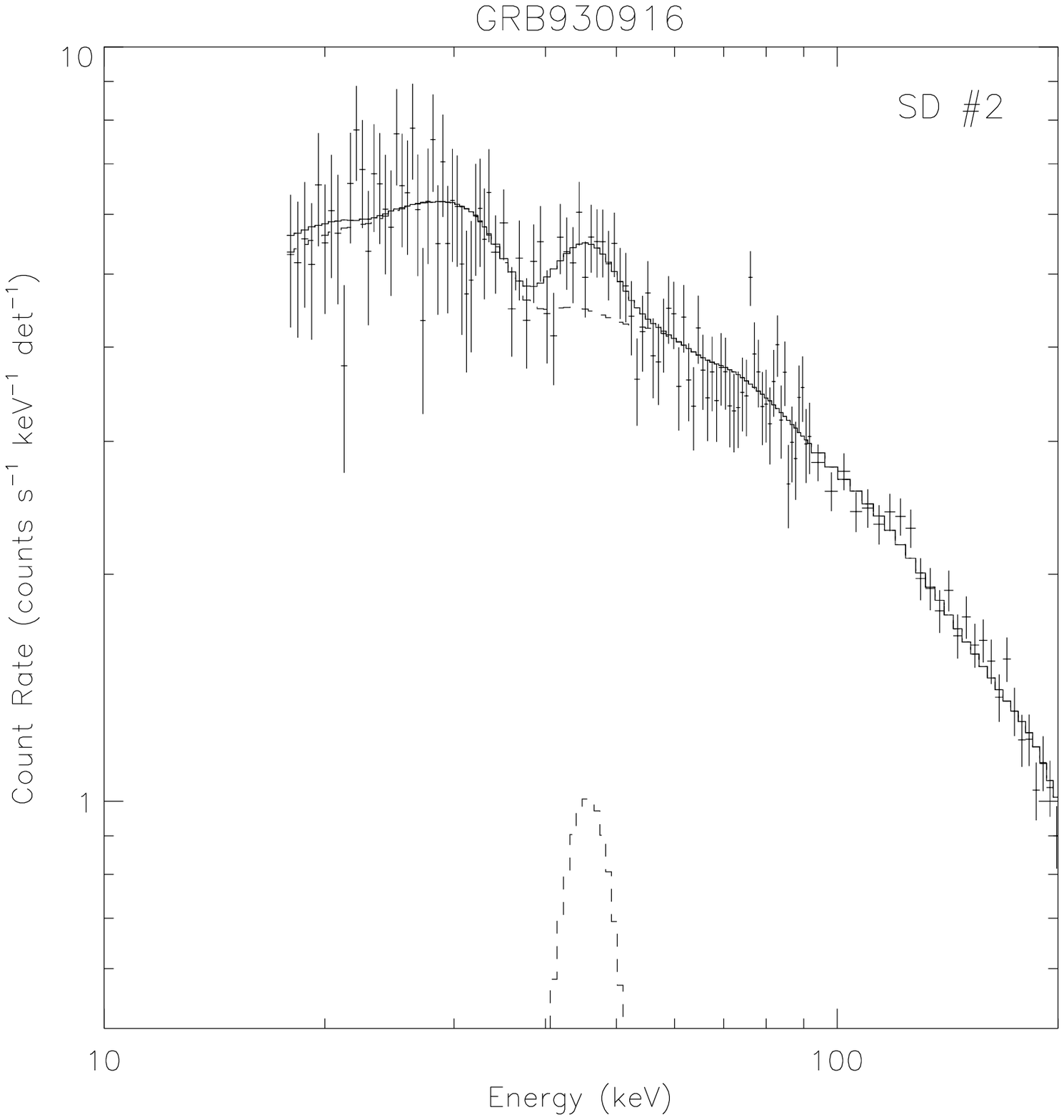,width=6.0cm,%
bbllx=55bp,bblly=185bp,bburx=550bp,bbury=674bp,clip=}
}
}
\caption{
FIGURE 1.  Data from the interval 22.144 to 83.200~s after the BATSE trigger
of GRB~930916. 
The plot shows the count rate data (points) and count rate models (histograms).
The count rate models are obtained by folding the photon models through
a model of the detector response.
The `bump' at 30 keV is expected from the K-edge of the iodine in NaI.
Left panel: best continuum-only fit to the data of SD~2.
The data show a clear excess above the model from 41 to 51 keV.
Right panel: A narrow spectral feature is added to the model: an emission line
at 45 keV improves $\chi^2$ by 23.1.   The width of the feature is due to the
detector resolution.   The solid histogram depicts the total count model; the
dashed histograms show the continuum and line portions separately.
}
\end{figure}

An interval of data from SD~2 for GRB~930916 (BATSE trigger 2533) which
contains the peak and trailing portion of the event
appears to have a significant line
($\Delta \chi^2 = 24.1$, $P = 6 \times 10^{-6}$).
Because of the coarser time resolution of the data from SD~7, 
a slightly different interval
must be used to compare the results from SD~2 and 7.   
Using the revised interval, the significance of the feature in SD~2
is slightly reduced to $\Delta \chi^2 = 23.1$ (Fig.~1).

There is no evidence for the feature in the data of SD~7 (Fig.~2).
Not only does adding a line to the model result in no 
improvement in $\chi^2$,
imposing a line at the strength expected according to SD~2 results in a
$\chi^2$ increase of 9.7 (Fig.~2, right).

\begin{figure}
\centerline{
\mbox{
\epsfig{file=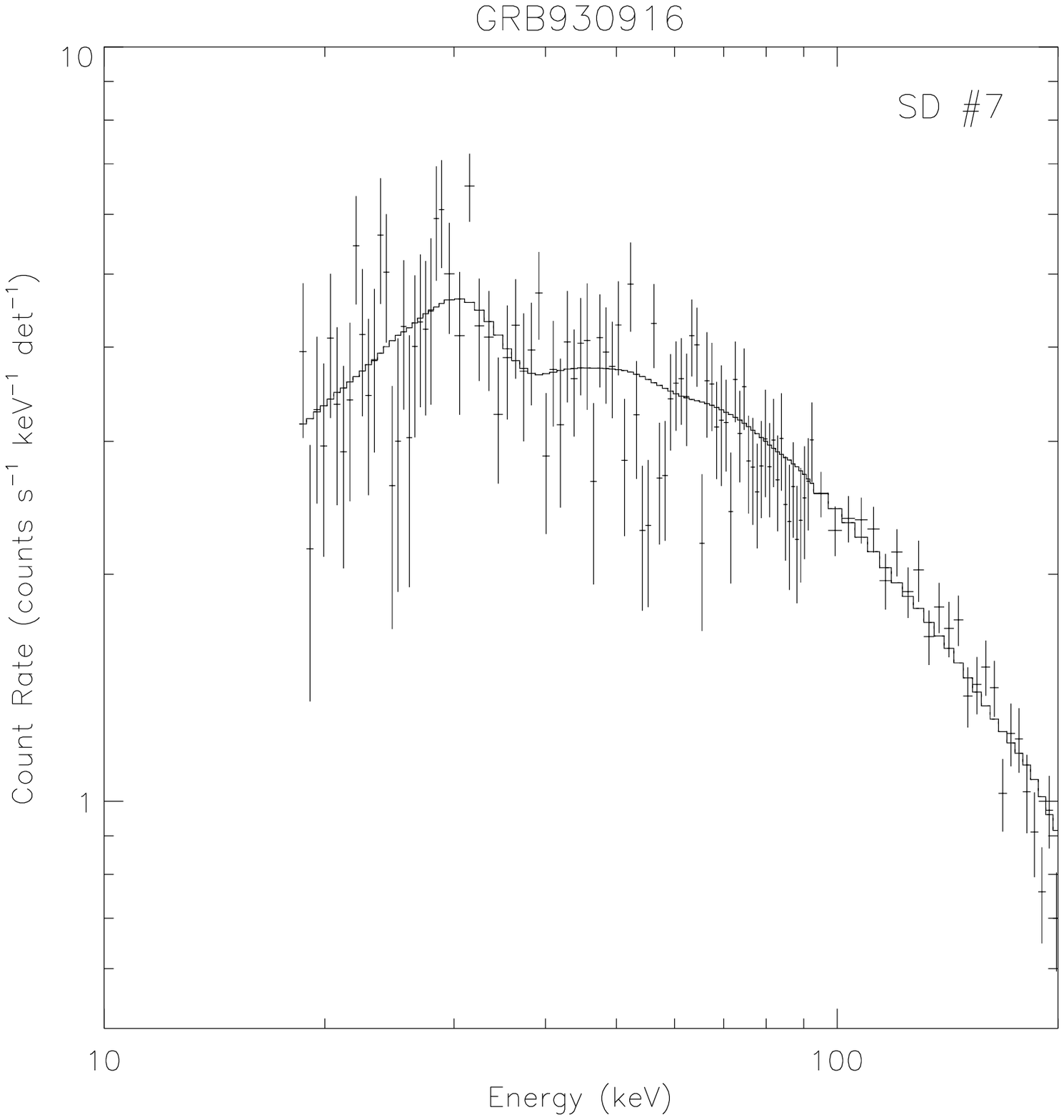,width=6.0cm,%
bbllx=55bp,bblly=185bp,bburx=550bp,bbury=674bp,clip=}
\psfig{file=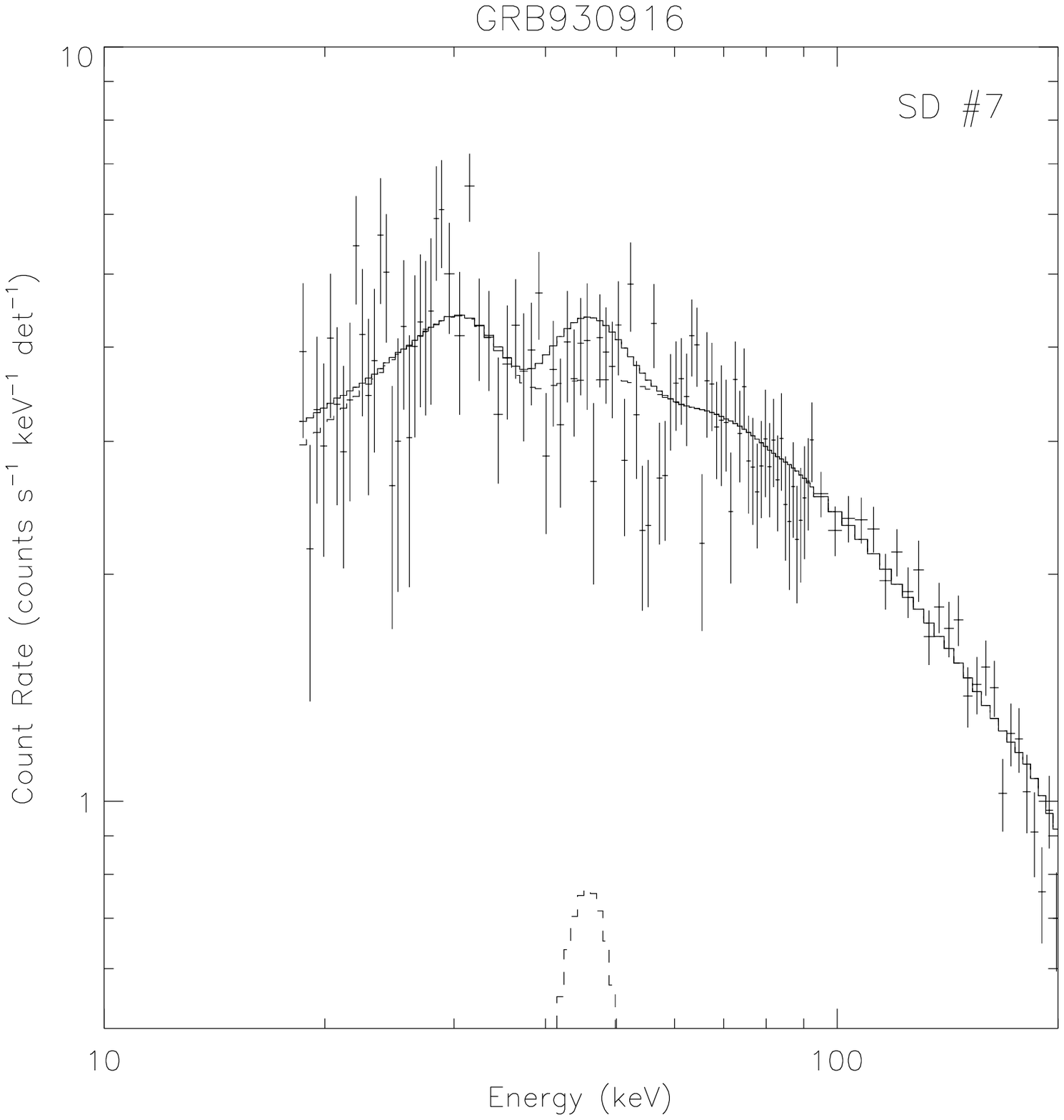,width=6.0cm,%
bbllx=55bp,bblly=185bp,bburx=550bp,bbury=674bp,clip=}
}
}
\caption{
FIGURE 2.  Left panel: best continuum-only fit to the data of SD~7 for
the same time interval of GRB~930916 as used in Fig.~1.
Adding a line results in no $\chi^2$ improvement.
Right panel: 
The continuum model is still a fit, but a line at the strength indicated
by the data of SD~2 is imposed.
The model in the region of the putative line is clearly above the data
and $\chi^2$ is increased (compared to the continuum-only fit of
the left panel) by
9.7, rather than decreased.
}
\end{figure}

A quantitative sensitivity or consistency analysis is required to decide
whether  the failure to detect the candidate in SD~7 is 
reasonable because of some difference between the detectors.
The two detectors have the same gain and viewing angles of
$31^\circ$ for SD~2 and
$64^\circ$ for SD~7.
Because of the detectors are almost as thick as their diameters, the
effective area has only a small dependence on burst angle.

We have performed simulations to quantitatively test the consistency of the two
detectors.  We use the joint fit to the data of the detectors that
viewed the burst as the best compromise between the line strengths preferred
from the data of each detector.   Then,
using the parameters of the joint fit photon model and the detector
response model, 
1000 simulated
count rate datasets are made for each detector.
These simulated spectra are fit to determine the range of line significances
expected.
A simulated significance above the observed  significance
for SD~2 ($\Delta \chi^2 = 23.1$) is obtained in 9\% of the simulations,
indicating that the observed significance is slightly better than expected.
However, a simulated significance $\Delta \chi^2$ below 0.1 is obtained in
only 2\% of the simulations of SD~7, indicating that the observation
is only marginally consistent with expectations.

\bsk
{\ni 3. CONCLUSIONS}
\ssk
\ni

Consistency between the results obtained from several detectors
is required for a believable result.
In the case of GRB~930916, the consistency is marginal.
The event could be understood as a 9\% probable fluctuation towards
high significance in SD~2 and a 2\% probable fluctuation towards insignificance
in SD~7.
If this were the only such case, such an explanation would be plausible.
There are several other such cases among the 13 candidates, raising the
possibility of a systematic error that invalidates the statistical
significances.
We have performed many tests of the reliability of the SDs [5]
and for systematics in the line analysis [4]: the SDs and the data pass
all tests.
Until we have a better understanding of these apparent inconsistencies 
between the
data collected from different detectors,
the reality of all of the BATSE line candidates is unclear.

A key lesson is the power of observations from more than one detector
for testing the reality of a possible line feature.
Agreement would be powerful confirmation; disagreement might indicate
a systematic error.

\bsk
\baselineskip = 12pt


{\references \ni REFERENCES
\ssk
%
\ref 1. Band, D. L., Ryder, S., Ford, et al.
1996, ApJ, 458, 746
%
%
\ref 2. Briggs, M. S., Band, D. L., et al.
1996, in Gamma-Ray Bursts, AIP Conf. Proc. 384, 153
%
\ref 3. Briggs, M. S., Band, D. L.,  et al.
1998, in Gamma-Ray Bursts, AIP Conf. Proc. 428, 299
%
\ref 4. Briggs, M. S., et al., 1998, ApJ, in preparation
%
\ref 5. Paciesas, W. S., Briggs, M. S., et al.
1996, in Gamma-Ray Bursts,
AIP Conf. Proc. 384, 213
}                      

\end{document}